\documentclass[preprint,showpacs,preprintnumbers,amsmath,amssymb,double-spaced]{revtex4}

\usepackage{graphicx}
\usepackage{dcolumn}
\usepackage{bm}

\begin{document}
\title{Entanglement in a Spin-$s$ Antiferromagnetic Heisenberg
Chain}

\author{Xiang Hao}

\author{Shiqun Zhu}
\altaffiliation{Corresponding author}

\email{szhu@suda.edu.cn}

\affiliation{School of Physical Science and Technology, Suzhou
University, Suzhou, Jiangsu 215006, People's Republic of China}

\begin{abstract}
The entanglement in a general Heisenberg antiferromagnetic chain
of arbitrary spin-$s$ is investigated. The entanglement is
witnessed by the thermal energy which equals to the minimum energy
of any separable state. There is a characteristic temperature
below that an entangled thermal state exists. The characteristic
temperature for thermal entanglement is increased with spin $s$.
When the total number of lattice is increased, the characteristic
temperature decreases and then approaches a constant. This effect
shows that the thermal entanglement can be detected in a real
solid state system of larger number of lattices for finite
temperature. The comparison of negativity and entanglement witness
is obtained from the separability of the unentangled states. It is
found that the thermal energy provides a sufficient condition for
the existence of the thermal entanglement in a spin-$s$
antiferromagnetic Heisenberg chain.
\end{abstract}

\pacs{03.65.Ud, 03.67.Mn, 75.10.Jm}

\maketitle

\section{Introduction}

The entanglement of quantum systems has been extensively
implemented to realize quantum computation and secure
communication. As an important resource in quantum information
processing \cite{Divin00}, it is necessary to qualify the
entanglement. The entanglement of formation \cite{Wootters98} and
the relative entropy of entanglement \cite{Bennett96,Vedral98} are
basic measures for the bipartite systems. Using these measures,
thermal entanglement \cite{Arnesen01,
Kamta02,Gu04,Subra04,Wang02,Wang022} has been investigated in some
solid state systems of Heisenberg spin-$\frac 12$ model.
Anisotropy effect \cite{Kamta02}, non-nearest interaction
\cite{Gu04}, high dimensions \cite{Subra04} and multiple qubits
\cite{Wang02} were considered. Meanwhile, the entanglement witness
\cite{Lewenstein00,Sanpera01,Dowling04,Toth05,Maruyama05,
Lunkes05,Hyllus05} for spin-$\frac 12$ systems was proposed. The
existence of entanglement was observed in an experimental
situation \cite{Ghosh03}. The thermal energy \cite{Dowling04,
Toth05} and the magnetic susceptibility \cite{Ghosh03} were
regarded as the entanglement witnesses for a macroscopic solid
state system. The effect of the edges of lattices \cite{Dowling04}
was considered. The entanglement of Bose-Hubbard model
\cite{Toth05} has been witnessed by the energy. Besides a
spin-$\frac 12$ model, a more universal quantum system focuses on
a high spin-$s$ Heisenberg model \cite{Haldane83,
Morra88,Mutka93,Granroth96,Jiang96,Wang99}. In the integer spin
systems such as $CsNiCl_3$ \cite{Morra88} and $MnCl_3(bipy)$
\cite{Wang99}, there is the exciting phenomenon of Haldane gap
\cite{Haldane83,Morra88,Mutka93,Granroth96,Jiang96, Wang99}.
Additionally, the efficiency of the quantum communication
\cite{Brukner02,Vaziri02,Bechmann00,Cerf02} was also enhanced by
utilizing the entanglement between two qutrits (a
three-dimensional quantum system). Due to many interesting
features of high-spin quantum systems, the entanglement in a
quantum Heisenberg system with arbitrary spin $s$ needs to be
studied. Recently, a computable measure of entanglement, i. e.,
the negativity \cite{Vidal02} has been theoretically generalized
to the high-spin systems using the separability principle
\cite{Peres96,Horodecki96}. Therefore, one entanglement witness
can be suggested to experimentally detect the entanglement in such
high-spin quantum systems.

In this paper, the entanglement in a spin-$s$ antiferromagnetic
Heisenberg chain is investigated. In Section II,  one entanglement
witness for high-spin quantum systems is introduced. Thermal
entanglement may be indicated by the characteristic temperature
where the thermal energy equals to the minimum energy of all
separable states. For bipartite lattices of spin-$s$, the analytic
expression of the minimum energy of the separable state is
deduced. In Section III, it is demonstrated that the thermal
energy provides a sufficient condition of the existence of the
thermal entanglement for high-spin systems compared to the
negativity.

\section{Entanglement witness for a spin-s Heisenberg chain}

For an isotropic spin-$s$ Heisenberg chain, the Hamiltonian $H$ is
given by,
\begin{equation}
\label{eq:1}
 H=\sum_{i=1}^LJ\vec{S}_i\cdot\vec{S}_{i+1}
\end{equation}
where $\vec{S}_i=(S_i^x,S_i^y,S_i^z)$ and
$S_i^{\alpha}(\alpha=x,y,z)$ are the spin-$s$ operators for the
{\it i}th spin, $J$ is the interaction coefficient. The spin
operators $S_i^x, S_i^y$ can be expressed by the lifting operator
and the lowering one, $S_i^{+}$ and $S_i^{-}$. In the Hilbert
space of $\{|m\rangle_i, m=-s, -s+1,...,s \}$,
$S_i^{\pm}|m\rangle_i=\sqrt {(s\pm m+1)(s\mp m)}|m\pm 1\rangle_i$
and $S_i^z|m\rangle_i=m|m\rangle_i$. The periodic boundary
condition of $L+1=1$ is assumed. The cases of $J>0$ and $J<0$
correspond to the antiferromagnetic and ferromagnetic cases
respectively. In the following discussion, an antiferromagnetic
chain is considered. The state at a thermal equilibrium
temperature $T$ is $\rho(T)=e^{-H/kT}/Z$ where $Z$ is the
partition function. For the convenience, both Boltzmann constant
$k$ and Planck constant $\hbar$ are assumed to be one. One
entanglement witness for a spin-$s$ quantum system can be
generalized to \cite{Dowling04, Toth05},
\begin{equation}
\label{eq:2}
W=\langle H \rangle-E_{min}
\end{equation}
where $\langle H \rangle=tr(\rho H)$ is the thermal energy at the
thermal state $\rho$ and $E_{min}$ is the minimum energy that any
separable state may be obtained. This minimum energy can always be
achieved by a pure separable state $|\psi\rangle_{sep}$. When the
value of $W$ is nonnegative, the state $\rho$ is the sparable
(unentangled) state. Only if $W<0$, there is the thermal
entanglement in the state of $\rho$. Because the ground energy
$E_0$ is always less than $\langle H \rangle$, there is a maximum
gap for entanglement, $G=|E_0-E_{min}|$. In Eq.~(\ref{eq:2}), the
solution of the minimum energy $E_{min}$ for any separable state
needs to be calculated. An isotropic spin-$s$ Heisenberg chain is
an example of bipartite lattices. The Hamiltonian can be written
by $H=\sum_{i=1}^{L}H_{i}$ where
$H_{i}=J\vec{S}_i\cdot\vec{S}_{i+1}$. If the minimum-energy
separable state $|\psi_i\rangle_{sep}$ for $H_{i}$ is known, the
total separable state for $H$ can be expressed by
$|\psi\rangle_{sep}=\prod_{i=1}^{L}|\psi_i\rangle_{sep}$. In the
case of an isotropic antiferromagentic chain, the state of
$|\psi_i\rangle_{sep}$ can be analyzed by the standard symmetry
methods~\cite{Eggeling01}. The minimum-energy separable state for
$H_i$ can be written as,
\begin{widetext}
\begin{eqnarray}
\label{eq:3} |\psi_i\rangle_{sep}=&&\frac 1{4^s}\sum_{m=0}^s\sqrt
{C_m}(|s-m\rangle_i+|m-s\rangle_i)\otimes\sum_{m=0}^s(-1)^m\sqrt
{C_m}(|s-m\rangle_{i+1}+|m-s\rangle_{i+1}) \mbox{,\quad 2s+1 is odd}\nonumber\\
|\psi_i\rangle_{sep}=&&\frac 1{4^s}\sum_{m=0}^{s-\frac 12}\sqrt
{C_m}(|s-m\rangle_i+|m-s\rangle_i)\otimes\sum_{m=0}^{s-\frac
12}(-1)^m\sqrt {C_m}(|s-m\rangle_{i+1}-|m-s\rangle_{i+1})
\mbox{,\quad 2s+1 is even}
\end{eqnarray}
\end{widetext}
When $2s+1$ is even, the coefficient satisfies $C_{m+1}=\frac
{2s-m}{m+1} C_m$. However, when $2s+1$ is odd, $C_{m+1}=\frac
{2s-m}{m+1} C_m$ for $m<s-1$ and $C_{s}=\frac {s+1}{4s}C_{s-1}$.
As an example, an antiferromagnetic Heisenberg chain with spin
$s=1$ is investigated. Without losing generality, the parameters
of the minimum-energy separable state $|A\rangle|B\rangle$ can be
assumed as,
\begin{equation}
\label{eq:4}
|j\rangle=a_j|1\rangle+b_je^{i\phi^1_j}|0\rangle+c_je^{i\phi^2_j}|-1\rangle
, \quad j=A,B
\end{equation}
By means of the standard symmetry method, $a_j=c_j$, $\phi^2_j=0$
and $\phi^1_A-\phi^1_B=\pi$.  To find the minimum energy, the
energy can be calculated by,
\begin{equation}
\label{eq:5} \langle A|\langle B|H|A
\rangle|B\rangle=-16Ja_j^2b_j^2, \quad (2a_j^2+b_j^2=1)
\end{equation}
It is easily seen that the minimum energy for any separable state
can be achieved by $a_j=\frac 12, b_j=\frac {\sqrt 2}2$.

For the simplest case of $L=2$, the ground state energy can be
expressed by $E_0=-2J(s^2+s)$ while the minimum energy for any
separable state is $E_{min}=-2Js^2$. Therefore, the maximum gap for
entanglement $G(s)$ is given by $G(s)=2Js$. The bigger gap is
obtained at the higher spin-$s$ system. That is, the entanglement is
easily detected in a high-spin system. There is a characteristic
temperature $T_c$ for $W=0$. Since $\langle H_i \rangle$ is
increased with increasing value of the temperature
\cite{Schliemann03}, it is evident that $W>0$ when $T>T_c$. It is
obviuos that the thermal entanglement between two nearest
neighboring spins exists only if $T<T_c$. In Fig. 1, the
characteristic temperature $T_c$ is plotted when the spin $s$ is
varied. It is found that $T_c$ is almost linearly increased with
$s$. The high spin quantum system can increase the temperature range
for the existence of the thermal entanglement.

For an $L$-partite Heisenberg chain, the corresponding minimum
energy is $E_{min}=-JLs^2$. There is also a characteristic
temperature $T_c$ below which the entanglement exists between
arbitrary two neighboring spins. The relation of $T_c$ to the total
number of lattices $L$ is shown in Fig. 2 where the coupling is
chosen to be $J=1$. The upper triangles represent the numerical
results of $T_c$ for spin $s=1$ while the lower squares denote the
values of $T_c$ for spin $s=\frac 12$. It is seen that the
characteristic temperatures $T_c$ for both different spin $s$ are
monotonously decreased with $L$ and then approaches a constant at
certain number of lattices. In the limit of $L\rightarrow\infty$,
the constant value for $s=1$ is approximately given by $T_c=1.05$
which is higher than that of $T_c=0.80$ for $s=\frac 12$. This is
consistent with recent analyses \cite{Wang02,Wang022,Toth05}. For
spin $s=\frac 12$, the constant value of the characteristic
temperature is $T_{cc}=0.8$ that is approximately $\frac 14$ of the
value in Ref. \cite{Toth05}. This is due to that the parameters
chosen in our numerical calculations are about $\frac 14$ of that in
Ref. \cite{Toth05}. When the number of lattices $L$ is very large,
it is very interesting to note that the difference $\Delta T_{cc}^s$
of the constant characteristic temperature $T_{cc}$ between
different spin $s$ is a function of $s$. That is, $\Delta
T_{cc}^s=T_{cc}^{s+\frac 12}-T_{cc}^s\thicksim 0.4s$ for $J=1$. The
fact that the characteristic temperature $T_c$ approaches a constant
can qualitatively explain the detection of the thermal entanglement
at finite temperature in a real solid state system of larger number
of lattices \cite{Ghosh03}.

\section{Relation of entanglement witness to negativity}

Through the thermal energy, the entanglement of a Heisenberg chain
can be witnessed. Based on the separability principle, The
negativity $N$ can be used to quantify the entanglement
\cite{Vidal02}. The negativity $N$ is introduced by,
\begin{equation}
\label{eq:6}
N(\rho)=|\sum_i\mu_i|
\end{equation}
where $\mu_i$ is the $i$th negative eigenvalue of $\rho^{T}$ which
is the partial transpose of the mixed state $\rho$. The measure
corresponds to the absolute value of the sum of negative
eigenvalues of $\rho^{T}$. For the separability of unentangled
states, the partial transpose matrix $\rho^{T}$ has nonnegative
eigenvalues if it is unentangled. As an example of thermal states
in an isotropic spin-$s$ antiferromagnetic chain, the relation of
entanglement witness to negativity is investigated.

Considering a two-spin isotropic antiferromagnetic Heisenberg
chain, any thermal state $\rho$ is an SU(2)-invariant state
\cite{Schliemann03}. In the case of $s=\frac 12$, the partial
transpose matrix $\rho^{T}$ has negative eigenvalues when the
correlation function satisfies ~\cite{Schliemann03},
\begin{equation}
\label{eq:7}
\langle \vec{S}_1\cdot \vec{S}_2 \rangle<-\frac 14
\end{equation}
For a thermal state, Eq.~(\ref{eq:7}) is also equivalent to
$\langle H \rangle<-\frac J2$ or $W<0$ where $E_{min}=-\frac J2$.
The negativity can also be expressed by,
\begin{equation}
N(\rho)=-\frac WJ
\end{equation}
It shows that the thermal entanglement exists for $N>0$ or $W<0$.
That is, both the entanglement witness and the negativity provides
the same condition for thermal entanglement in the case of
$s=\frac 12$. The temperature range for thermal entanglement is
given by $T<\frac {2J}{\ln 3}$. However, for a thermal state of
$s=1$, the negative partial transpose needs,
\begin{equation}
\label{eq:9} \langle (\vec{S}_1\cdot \vec{S}_2)^2 \rangle>2
\end{equation}
which is also expressed by $\langle H^2 \rangle>8J^2$.
Eq.~(\ref{eq:9}) determines a temperature range for the existence
of the entanglement. That is, the entanglement exists when
$T<\frac {2J}{\ln 2.08}$. Compared with the entanglement witness
of Eq.~(\ref{eq:2}), the thermal energy satisfies,
\begin{equation}
\label{eq:10} \langle H \rangle<-2J
\end{equation}
This temperature range of Eq.~(\ref{eq:10}) is $T<\frac {6J}{\ln
10}$. It shows that the area of thermal entanglement decided by
the negativity is larger than that determined by the entanglement
witness. The exact relation of negativity and entanglement witness
can be expressed as,
\begin{equation}
N(\rho)=\frac 1{8J^2}[(W-2J)^2+V(H)]-1
\end{equation}
where the variance $V(H)$ is written by $V(H)=\langle H^2
\rangle-\langle H \rangle^2$. When the temperature $T\geq T_c$, the
entanglement witness may be assumed to $W=0$. The difference of
$\Delta=N-|W|$ is plotted in Fig. 3 when the temperature and
coupling are varied. It shows that there is almost no differences
for the weak coupling in Fig. 3(a). When the coupling $J$ is
increased, the difference becomes large. The contour map is shown in
Fig. 3(b) where the dotted line represents $W=0$. Since the
temperature area of entanglement decided by negativity is larger
than that by the witness, the difference $\Delta=0$ corresponds to
the negativity $N=0$. It is seen that the critical temperature of
$N$ is higher than that of $W$. It demonstrates that the
entanglement witness $W$ provides a more sufficient condition for
thermal entanglement.

\section{Discussion}

The entanglement in an isotropic spin-$s$ antiferromagnetic
Heisenberg chain is investigated using the entanglement witness of
thermal energy and the negativity. The analytic expression of the
minimum-energy separable state is deduced. The entanglement
witness determines a characteristic temperature $T_c$ below which
an entangled thermal state can be obtained. It is found that the
characteristic temperature is almost linearly increased with the
increasing number of spin $s$. For an $L$-partite spin chain,
$T_c$ decreases with increasing the number of lattices. However,
$T_c$ approaches a constant when the number of lattices is very
large. This shows that the entanglement can be detected in a real
solid state system of large number of lattices even for finite
temperature. It is also shown that the characteristic temperature
is a linear function of the coupling. From the separability
principle, the entanglement witness is different from the
negativity in detecting thermal entanglement of high-spin quantum
systems. The thermal energy provides a more sufficient condition
for the existence of the entanglement.

\begin{acknowledgements}
It is a pleasure to thank Yinsheng Ling, Jianxing Fang, and Qing
Jiang for their many fruitful discussions about the topic.
\end{acknowledgements}

\bibliography{XHao6}

\newpage

{\Large {\bf Figure Caption}}

\vskip 0.6cm

{\large Fig. 1}

The characteristic temperature $T_c$ is plotted when the spin $s$ is
varied.

{\large Fig. 2}

The characteristic temperature $T_c$ is plotted as a function of the
total number of lattices $L$. The upper triangles are the results of
$T_c$ for $s=1$. The corresponding constant value is about
$T_c=1.05$. The lower squares represent the values of $T_c$ for
$s=\frac 12$, and the constant value of $T_c$ is $0.80$.

{\large Fig. 3}

(a). The difference $\Delta=N-|W|$ of the negativity $N$ and the
witness $W$ is plotted when the temperature and coupling are varied;

(b). The corresponding contour map. The dotted line represents
$W=0$.

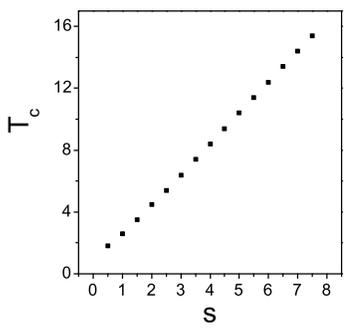
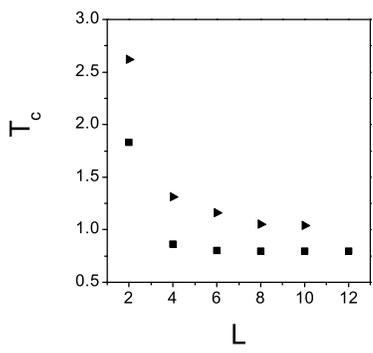
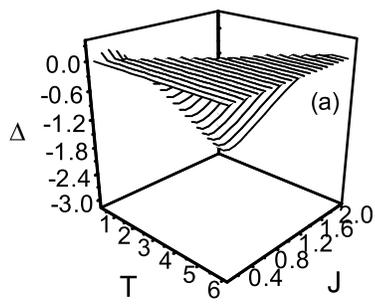
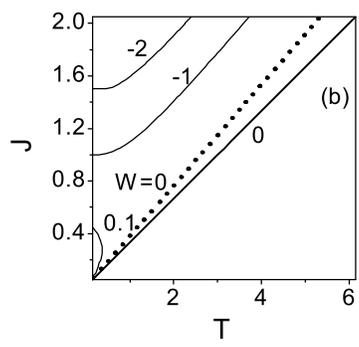
\end{document}